\documentclass[10pt,twocolumn,preprintnumbers,nofootinbib,prd,superscriptaddress,aps]{revtex4-1}

\usepackage[utf8]{inputenc} 
\usepackage{graphicx,amssymb,amsmath,amsthm,amsfonts,epstopdf,epsfig,epsf,times}
\usepackage[linktocpage]{hyperref}
\usepackage[usenames]{color}
\usepackage{epstopdf}
\usepackage{textcomp}
\usepackage{bm}
\usepackage{latexsym}
\usepackage{rotating}
\usepackage{hyperref}
\usepackage{color}
\usepackage{longtable}
\usepackage{enumerate}
\usepackage{tensor}
\usepackage{stmaryrd}
\usepackage[normalem]{ulem}
\usepackage{mathtools}
\usepackage{url}
\usepackage{multirow}
\usepackage{graphicx}
\usepackage[caption=false]{subfig}
\usepackage{mathtools}
\usepackage{verbatim}
\usepackage{soul,xcolor}
\usepackage{xspace}
\usepackage{amsmath}
\usepackage{xfrac}
\setstcolor{red}
\usepackage{esint}
\usepackage{physics}

\makeatletter
\DeclareRobustCommand\onedot{\futurelet\@let@token\@onedot}
\def\@onedot{\ifx\@let@token.\else.\null\fi\xspace}
 
\makeatother

\definecolor{coolblack}{rgb}{0.0, 0.18, 0.39}
\definecolor{darkred}{rgb}{0.5,0,0}
\definecolor{darkgreen}{rgb}{0,0.5,0}
\definecolor{darkblue}{rgb}{0,0,0.5}
\definecolor{lapislazuli}{rgb}{0.15, 0.38, 0.61}
\definecolor{venetianred}{rgb}{0.78, 0.03, 0.08}
\definecolor{bleudefrance}{rgb}{0.19, 0.55, 0.91}
\definecolor{dogwoodrose}{rgb}{0.84, 0.09, 0.41}
\definecolor{dogwoodrose}{rgb}{0.84, 0.09, 0.41}
\definecolor{darkorgane}{rgb}{1,0.549,0}
\hypersetup{colorlinks=true, citecolor=darkblue, linkcolor=darkblue, 
urlcolor = darkblue}
\definecolor{olive}{rgb}{0.5, 0.5, 0.0}

\renewcommand{\vec}[1]{\boldsymbol{#1}}

\newcommand{\ben}{\begin{enumerate}}
\newcommand{\een}{\end{enumerate}}
\newcommand{\Ord}[2]{\mathcal O \left(#1^{#2}\right)}

\def\be{\begin{equation}}
\def\ee{\end{equation}}
\newcommand{\beq}{\begin{eqnarray}}
\newcommand{\eeq}{\end{eqnarray}} 
\newcommand{\ba}{\begin{align}}
\newcommand{\ea}{\end{align}}

\newcommand{\submu}{_{\mu}}

\newcommand{\submunu}{_{\mu \nu}}
\newcommand{\supmu}{^{\mu}}
\newcommand{\supnu}{^{\nu}}

\newcommand{\mB}{\mathcal{B}}

\newcommand{\lr}[1]{\left( #1 \right)}

\def\be{\begin{equation}}
\def\ee{\end{equation}}

\begin{document}

\title{Electromagnetic radiation reaction and energy extraction from black holes: \\ The tail term cannot be ignored}

\author{João S. Santos}
\affiliation{CENTRA, Departamento de F\'{\i}sica, Instituto Superior T\'ecnico -- IST, Universidade de Lisboa -- UL,
Avenida Rovisco Pais 1, 1049-001 Lisboa, Portugal}
\author{Vitor Cardoso}
\affiliation{CENTRA, Departamento de F\'{\i}sica, Instituto Superior T\'ecnico -- IST, Universidade de Lisboa -- UL,
Avenida Rovisco Pais 1, 1049-001 Lisboa, Portugal}
\affiliation{Niels Bohr International Academy, Niels Bohr Institute, Blegdamsvej 17, 2100 Copenhagen, Denmark}
\author{José Natário}
\affiliation{CAMSDG, Departamento de Matem\'{a}tica, Instituto Superior T\'ecnico -- IST, Universidade de Lisboa -- UL,
Avenida Rovisco Pais 1, 1049-001 Lisboa, Portugal}
\begin{abstract} 
We study electromagnetic radiation reaction in curved space and the dynamics of radiating charged particles. The equation of motion for such particles is the DeWitt-Brehme equation, and it contains a particularly complicated, non-local, tail term. It has been claimed that the tail term can be neglected in certain magnetized black hole spacetimes, and that radiation reaction may then lead to energy extraction (``orbital widening") in the absence of an ergoregion. 
 We show that such claims are incorrect, at least in the Newtonian limit: the tail term can never be neglected consistently in the relevant scenarios, and when it is included the reported energy extraction no longer occurs. Thus, previous results are called into question by our work.
%
%
\end{abstract}

\maketitle

\section{Introduction\label{sec:introduction}}
Black holes (BHs) are the most compact objects in the cosmos, and matter in their vicinity can attain velocities close to the speed of light. They are therefore good particle accelerators and can efficiently convert gravitational energy to electromagnetic (EM) radiation~\cite{Ruffini:1973pc}. BHs are also prone to energy extraction mechanisms, whereby their rotational energy can be converted onto kinetic energy of particles or occupation number of fields~\cite{Penrose:1969pc,PhysRevD.12.2959,1977MNRAS.179..433B,Brito:2015oca}. This is a particularly appealing explanation for possible production of ultra high energy cosmic rays, with energies that can surpass $10^{20}$ eV \cite{deMelloNeto:2020xfa,TelescopeArray:2021zox}, or for other violent astrophysical phenomena. 
However, mechanisms for energy extraction, relying on rotation alone (either via the original Penrose process or variants thereof) are not efficient enough to explain the origin of cosmic rays or extremely energetic phenomena~\cite{1974ApJ...191..231W,Bardeen:1972fi,1975ApJ...196L.107P,PhysRevD.16.1615,PhysRevLett.113.261102,PhysRevLett.114.251103}.

Alternatives to the above simple scenario include the addition of external magnetic fields, which can increase the
efficiency of energy extraction via Penrose-like mechanisms~
\cite{1985ApJ...290...12W,1986ApJ...301.1018W,1986ApJ...307...38P,Tursunov:2019oiq,Kolos:2020gdc,2021Univ....7..416S,Tursunov:2020juz,Gupta:2021vww}. 
Magnetic fields can be supported by accretions disks, and there is ample evidence that these are u\-bi\-qui\-tous around stellar-mass and supermassive BHs~\cite{piotrovich_magnetic_2011,Eatough:2013nva,Baczko:2016opl,Daly:2019srb}.
Since asymptotically flat solutions of the field equations in the presence of magnetic fields are difficult to obtain, it is standard to assume that magnetic fields are of small enough amplitude so that they can be considered as test fields on a fixed background, such as the Schwarzschild geometry. For this to be a good approximation, the field should decay sufficiently fast far away from the BH, and its backreaction should be negligible. The last condition imposes the constraint~\cite{Galtsov:1978ag} 
\begin{equation}
	B \ll B_G = 2.4\times10^{19} \left( \frac{M_\odot}{M}\right) \quad \text{Gauss}\, ,
\end{equation}
where $M_\odot$ is the mass of the sun. Typical amplitudes around stellar mass and supermassive BHs are of order $10^8-10^4$ Gauss, and so there is ample room to be in the test-field approximation~\cite{piotrovich_magnetic_2011,Eatough:2013nva,Baczko:2016opl,Daly:2019srb}.

A characterisation of the circular orbits of charged particles around magnetized BHs was obtained e.g.\ in Ref.~\cite{Kolos:2015iva}, neglecting radiation emission.
Emission of radiation changes the orbit of the particle itself, making the problem of solving for its motion highly nontrivial.

Here, we discuss radiation reaction in general relativity (GR), in the context of charged particle motion. It was claimed in Ref.~\cite{2018ApJ...861....2T} that self-force effects in magnetized BH spacetimes lead to some charged particles {\it gaining} energy (``orbital widening") in the absence of an ergoregion. This is a highly counter-intuitive result -- and in the present work we show that it is incorrect, at least in the Newtonian limit.

We use a system of units with $G=c=1$, and use Gaussian units for electromagnetism, meaning that we have $4\pi\varepsilon_0 = 1$ and $\mu_0 = 4 \pi$. The metric signature is $\left( -,+,+,+ \right)$, greek indices run from 0 to 3, and boldface quantities are three-vectors.
\section{Radiation reaction force in curved space\label{sec:rad}}
The equation of motion of a pointlike particle of charge $q$ and mass $m$ in curved space was studied by DeWitt and Brehme \cite{DEWITT1960220}, with an important correction by Hobbs~\cite{HOBBS1968141}. In the presence of radiation reaction, it reads
\beq
\frac{D u\supmu}{d \tau}&=&\frac{q}{m} F^{\mu}_{\ \ \nu} u^{\nu} + \frac{2 q^2}{3 m} \left(\frac{D^2 u^{\mu}}{d \tau^2} + u^\mu u_\nu \frac{D^2 u^\nu}{d \tau^2}\right) \nonumber\\
&+& \frac{q^2}{3 m} \left(R^{\mu}_{\ \ \nu} u^\nu + R^{\nu}_{\ \ \alpha} u_\nu u^\alpha u^\mu\right)+ \frac{2 q^2}{m} f^{\mu \nu}_{\rm tail} u_\nu\,,\label{eq:rad_DeWitt_Brehme}
\eeq
where $u^\mu (\tau)$ is the particle's four-velocity, $\tau$ is the proper time, $F\submunu = \partial_\mu A_\nu - \partial_\nu A_\mu$ is the Faraday tensor with EM four-potential $A_\mu$, and $R\submunu$ is the Ricci tensor (identically zero for vacuum spacetimes). 
The first two terms on the RHS of Eq.~\eqref{eq:rad_DeWitt_Brehme} were already present in flat space \cite{Jackson:1998nia,Poisson:1999tv}, and correspond, respectively, to the Lorentz and Abraham-Lorentz-Dirac force. The last two terms are curved spacetime effects: a coupling to the Ricci curvature, and the so-called tail term, given by
\begin{equation}
	f^{\mu \nu}_\text{tail} = \int _{-\infty} ^{\tau^-} \nabla^{[ \mu} G^{\nu]}_{\ \ \lambda^\prime} \lr{z(\tau), z(\tau^\prime)} u^{\lambda^\prime} d \tau^\prime \, ,
	\label{eq:rad_tail}
\end{equation}
where square brackets denote anti-symmetrisation, $z(\tau)$ is the world line of the particle, $G^{\nu}_{\ \ \lambda^\prime} \lr{x,x^\prime}$ is the retarded Green's function for the vector wave equation in curved space, and the integral is taken over the entire history of the particle. For a modern derivation of Eqs.~\eqref{eq:rad_DeWitt_Brehme} and \eqref{eq:rad_tail} see the review by Poisson \cite{poisson_motion_2004}.

Equation~\eqref{eq:rad_DeWitt_Brehme} is of third order in the position of the particle, and so it has a number of unwanted features, such as the existence of runaway solutions \cite{poisson_motion_2004}. To circumvent this issue, order reduction of the original equation can be achieved, using a procedure originally introduced by Landau and Lifshitz ~\cite{Landau:1975pou}. This procedure assumes that the radiative terms are small in comparison with the other terms, so that
\begin{equation}
	\frac{D^2 u\supmu}{d \tau^2} \approx \frac{q}{m} \frac{D}{d \tau} \lr{F\supmu _{\ \ \nu} u\supnu}\, .
 \label{eq:approx}
\end{equation}
The validity of this approximation is discussed elsewhere~ \cite{poisson_motion_2004,2001PhLA..283..276R,Spohn:1999uf}. Replacing Eq.~\eqref{eq:approx} in Eq.~\eqref{eq:rad_DeWitt_Brehme} we obtain the reduced equation of motion, which in vacuum spacetimes ($R\submunu = 0$) takes the form
\beq
\frac{D u\supmu}{d\tau} & =&\frac{q}{m} F\supmu _{\ \ \nu} u\supnu + \frac{2 q^2}{3 m} \left(\frac{q}{m}\nabla_\alpha F^\mu _{\ \ \nu} u^\alpha u^\nu\right) \nonumber\\
&+& \frac{2 q^2}{3 m} \left(\frac{q^2}{m^2} \left(F^{\mu \nu}F_{\nu \rho} + F^{\nu \alpha}F_{\alpha \rho} u_\nu u^\mu \right)u^\rho \right)\nonumber\\
& +& \frac{2 q^2}{ m}f^{\mu \nu}_\text{tail} u_\nu \,.
\label{eq:rad_motion_reduced}
\eeq
Here, the only new term with respect to the reduced Abraham-Lorentz-Dirac equation in flat space is the tail term. Since it involves an integral over the entire history of the particle, its computation is challenging. Nevertheless, we show below that, in the Newtonian limit, all problems of dynamics of radiating charged particles in curved vacuum spacetimes fall into one of three classes: (i) radiation can be ignored altogether; (ii) space is so weakly curved that the flat space description of the system is valid; or (iii) the tail term {\em must} be included.

\section{Motion of particles around magnetized Schwarzschild black holes\label{sec:mot}	}

\subsection{Motion without radiation\label{sec:mot_no_rad}	}
We want to study a Schwarzschild BH immersed in an asymptotically uniform test magnetic field, $\vec{B} \to B_0 \vec{\hat{z}}$ at spatial infinity (where $\vec{\hat{z}}$ is the unit vector vector along the asymptotic $z$-axis). Therefore the spacetime metric is just that of the usual Schwarzschild solution \cite{Schwarzschild:1916uq}, given in the standard spherical coordinates $\lr{t,r,\theta,\phi}$ by
\beq
g &=& - f(r) dt^2 +\frac{dr^2}{f(r)}+ r^2 d\theta^2 + r^2 \sin^2 \theta \; d\phi^2 \, ,
\label{eq:mot_metric} \\
f(r) &=& 1 - \frac{2 M}{r}\, ,
\eeq
where $M$ is the BH mass. The magnetic field can be constructed by using the results of Ref.~\cite{PhysRevD.10.1680}, yielding
\begin{equation}
\begin{split}
F\submunu & = - B_0 \nabla_\nu Y_\mu \\
& = 2 B_0 \lr{r \sin^2 \theta \delta^r _{[\mu} \delta^\phi _{\nu]} + r^2 \sin \theta \cos \theta \delta^\theta _{[\mu} \delta^\phi _{\nu]} }\, ,
\end{split}
\end{equation}
where $Y\supmu = \delta ^\mu _\phi$ is the axial Killing vector field. This EM field corresponds to the vector potential
\begin{equation}
A\submu = \frac{B_0}{2} Y\submu = \frac{B_0}{2} r^2 \sin^2 \theta \delta \submu ^\phi \, .
\end{equation}
We can use Eq.~\eqref{eq:rad_motion_reduced} to study the orbits of charged particles with charge $q$ and mass $m$. As a first approximation, we ignore the radiative terms $\propto q^2$. Define the useful magnetic parameter
\begin{equation}
\mB = \frac{q B}{2 m},
\end{equation}
as well as the usual conserved quantities associated with the Killing vector fields  $X\supmu = \delta^\mu_t$ and $Y\supmu = \delta^\mu_\phi$:
\beq
E &=& - \lr{u_\mu + \frac{q}{m} A_\mu} X^\mu = f(r) u^t \, ,\\
L &=& \lr{u_\mu + \frac{q}{m} A_\mu}Y^\mu = r^2 \sin^2 \theta \lr{u^\phi + \mB}\, .
\eeq
Here $E$ is the energy per unit mass and $L$ is the angular momentum along the $\vec{\hat{z}}$ direction per unit mass.
By employing the Hamiltonian formalism \cite{Kolos:2015iva}, we find that, for $\mB\neq0$, circular orbits exist only in the equatorial plane and satisfy the condition 
\begin{equation}
E^2 = V_{\text{eff}}(r) = f(r) \left[ 1 + \left( \frac{L}{r} - \mB r \right)^2 \right]\, . 
\end{equation}
Note that the effective potential exhibits the symmetry $\lr{\mB, L} \to \lr{-\mB, -L}$. This feature allows us to distinguish between two different types of orbits:
\begin{itemize}
\item \emph{minus configuration} ($L<0$, $\mB>0$): this is the usual configuration, where the Lorentz force is attractive.
\item \emph{plus configuration} ($L>0$, $\mB>0$): this configuration cannot happen in flat space, and is characterized by a repulsive Lorentz force.
\end{itemize}
%
\subsection{Newtonian limit of motion with radiation\label{sec:mot_rad}}
We now aim to introduce radiation reaction into the equation of motion. We focus on the Newtonian limit (weak field and slow motion), since an approximate analytic expression for the tail term, due to DeWitt and DeWitt \cite{DeWitt:1964de}, is available in this case (which suffices to make our point).

To obtain the Newtonian limit, we start by recalling the expression for the gravitational potential of a pointlike particle of mass $M$ placed at the origin:
\begin{equation}
\Phi(r) = - \frac{M}{r}\, .
\end{equation}
Let $u\supmu$ be the 4-velocity of the charged test particle:
\begin{equation}
    u^\mu = u^t\lr{1,\vec{v}} \, , \quad u^t = \frac{d t}{d \tau} \sim \lr{1-v^2}^{-\frac{1}{2}} \, ,
	\label{eq:mot_ut}
\end{equation}
where $\vec{v}$ is the coordinate 3-velocity and $v$ is its norm. We now must impose that the particle is moving slowly through a weak gravitational field, which can be written as
\begin{equation}
	\left| \Phi \right| \sim v^2 \sim r^2 \mB^2 \sim \Ord{\varepsilon}{} \ll 1 \, .
	\label{eq:mot_assumptions}
\end{equation}
The first condition implies a weak gravitational field, the second condition states that the particle is slow-moving, and the third condition means that the  magnetic field is also weak\footnote{From the definitions \eqref{eq:mot_compare_dissipative} of the typical timescales of the problem, this condition may also be seen as stating that all the typical velocities of the problem are much smaller than the speed of light.}. From this point onward we neglect terms which are $\Ord{\varepsilon}{2}$ or smaller. The metric \eqref{eq:mot_metric} is then approximated by
\be
g_\text{Newton} = - \lr{1 + 2 \Phi} dt^2 + \lr{1-2\Phi}  d\vec{r}^2 \, .
\ee
If we calculate the action for timelike geodesics using this metric, we retrieve the appropriate Newtonian action
\be
S = \int \sqrt{- g\submunu u\supmu u\supnu} d\tau = \int \lr{1-\frac{v^2}{2}-\frac{M}{r}}d t\,,\label{eq:mot_newtonian_limit}
\ee
up to $\Ord{\varepsilon}{2}$ terms.
Now consider a particle in a circular orbit in the equatorial plane with radial coordinate $R$. Note that we will parameterize the motion using the coordinate time $t$ rather than the proper time $\tau$, since the $\Ord{\varepsilon}{}$ term in \eqref{eq:mot_ut} always leads to $\Ord{\varepsilon}{2}$ terms in the equation of motion. With that in mind, we use $u^t\approx 1$ from here on, consistently with the pre-relativistic idea that proper time is the same as coordinate time. We assume the effects of radiation are small enough for the particle to always be in a circular orbit, but that the radius $R$ of the latter changes adiabatically. In that case the position $\vec{x}(t)$ and the 3-velocity, $\vec{v}(t)$ of the charged particle are
\beq
\vec{x}(t) &=& R\lr{ \cos(\omega t)\, , \,  \sin(\omega t)\, , \, 0}\, ,\\
\vec{v}(t) &=& \frac{d\vec{x}(t)}{d t} \approx R\omega \lr{-\sin(\omega t)\, , \, \cos(\omega t)\, , \, 0} \, ,
\eeq
where the angular velocity $\omega$ also changes adiabatically.
We assume that the orbit satisfies all the conditions for the Newtonian limit to hold. In that case, the limit of Eq.~\eqref{eq:rad_motion_reduced} for these circular orbits is easy to find:
\begin{equation} \label{eq:mot_newtonian_eq}
	m \frac{d \vec{v}}{d t} = \vec{F}_\text{N} + \vec{F}_\text{L} + \vec{F}_\text{RR} + \vec{F}_\text{tail}\, . 
\end{equation}
Here, the first two forces on the RHS are, respectively, the Newtonian gravitational force and the Lorentz force, which, in this case, take the form
\begin{equation}
	\vec{F}_\text{N} = - m \frac{M}{R^2} \vec{\hat{r}} \, , \quad
	\vec{F}_\text{L} = q v B_0 \vec{\hat{r}} \, ,
\end{equation}
where $\vec{\hat{r}}$ is the unit radial vector. The third term comes from the second and third terms on the right-hand side of Eq.~\eqref{eq:rad_motion_reduced}. For the particle orbits we are studying, it is
\begin{equation}
\vec{F}_\text{RR} = \frac{2 q^3}{3m} B_0 \frac{v^2}{R} \vec{\hat{\phi}}\, ,\label{eq:mot_fforce_rr}
\end{equation}
where $\vec{\hat{\phi}}$ is the unit tangential vector. This is the first dissipative term we encounter. It is parallel to the velocity in plus configuration orbits, hence doing positive work, and anti-parallel to the velocity in minus configuration orbits, where it does negative work. Finally, we have the tail term, which was estimated for the weak field, slow motion regime in Refs.~\cite{Pfenning:2000zf,DeWitt:1964de}:
\begin{equation}
	\begin{split}
		\vec{F}_\text{tail} 
		&= q^2 \frac{M}{R^3} \vec{\hat{r}} - \frac{2 q^2}{3} \frac{d}{dt} \lr{\vec{\nabla}\Phi} \\
		&= q^2 \frac{M}{R^3} \vec{\hat{r}} - \frac{2 q^2}{3} \frac{M}{R^3} v\vec{\hat{\phi}} \, ,
	\end{split}
	\label{eq:mot_force_tail}
\end{equation}
where $\vec{\nabla}$ is the spatial coordinate gradient and $v\vec{\hat{\phi}}=\vec{v}$, that is, we are taking $v$ to have a sign. We can distinguish two components of the tail: a conservative part, $F^r _\text{tail}$ , and a dissipative part, $F^\phi _\text{tail}$ . We will now determine whether or not these tail terms must be included to adequately describe the motion of the charged particle.

The conservative part, $F^r _\text{tail}$, was originally derived in Ref.~\cite{DeWitt:1964de} (see also Ref.~\cite{Smith:1980tv}), and can be compared to the gravitational force as it points in the radial direction. If this comparison is done at the BH horizon $\lr{r=2 M }$ we obtain
\begin{equation}
	\frac{F^r _\text{tail}}{F_\text{N}} \approx 10^{-19} \lr{\frac{q}{e}}^2 \lr{\frac{m_e	}{m}} \lr{\frac{10 M_\odot}{M}}\, ,
	\label{eq:mot_compare_conservative}
\end{equation}
where $e$ and $m_e$ are, respectively, the electron charge and mass, and $M_\odot$ is the mass of the sun. This ratio is very small for astrophysically relevant scenarios, which means the conservative part of the tail term can be neglected. Note that it may seem odd that the comparison is performed at the BH horizon, since a weak field approximation is assumed; however, the tail term decays faster with $r$ than $F_\text{N}$, and so this is, in fact, an upper bound.

Regarding the dissipative part, $F^\phi _\text{tail}$, we start by noting it always does negative work, since it points opposite to the particle's velocity. To see if it must be included, we should compare it with the other dissipative term, $F _\text{RR}$, which is also tangent to the particle's velocity. We find
\begin{equation}
	 \frac{F_\text{RR}}{F_\text{tail} ^\phi} = \frac{\omega_c \omega}{\omega _K ^2} = \frac{F_\text{L}}{F_\text{N}} \, ,
	\label{eq:mot_compare_dissipative}
\end{equation}
(where $\omega = d \phi/d t$ is the orbital frequency, $\omega_c = 2 \mB$ is the cyclotron frequency, and $\omega_k= \sqrt{\frac{M}{R^3}}$ is the Keplerian frequency).  Thus, neglecting the dissipative part of the tail term and keeping the radiation reaction due to the Lorentz force is equivalent to neglecting the gravitational force while keeping the Lorentz force.

For vacuum spacetimes, therefore, the conservative part of the tail term can be neglected in the Newtonian limit.  However, the dissipative part can only be neglected in a consistent way if either of the following is true: 

\noindent {\bf (i)} Radiation reaction due to the Lorentz force is neglected as well, and the equation of motion is just Eq.~\eqref{eq:rad_DeWitt_Brehme} with all terms $\propto q^2$ set to zero;

\noindent {\bf (ii)} the gravitational force is neglected as well, and a flat-space description using the Abraham-Lorentz-Dirac equation is valid~\cite{Jackson:1998nia}.

In other words, in situations where the gravitational force dominates, such as the plus configurations orbits, the tail term \emph{must} be included, since $F_\text{tail} ^\phi$ also dominates $F_\text{RR}$. In particular, the particle will \emph{always} loose energy. This is the main result of this work.

As a sanity check, let us recall the non-relativistic expression for the radiation reaction force, \textit{i.e.}, the Lorentz self-force \cite{Jackson:1998nia}:
\begin{equation}
	\vec{F}_\text{rad} = \frac{2 q^2}{3} \frac{d \vec{a}}{d t} \,,
	\label{eq:mot_lorentz}
\end{equation}
where $\vec{a} = d \vec{v}/d t$. It is easy to show that the expressions for $\vec{F}_\text{RR}$ and $\vec{F}_\text{tail} ^\phi$ in Eqs.~\eqref{eq:mot_fforce_rr} -- \eqref{eq:mot_force_tail} are exactly what we would get by replacing the accelerations caused by the Lorentz and gravitational forces, respectively, in Eq.~\eqref{eq:mot_lorentz}. A further sanity check consists in verifying that the Newtonian limit of Eq.~\eqref{eq:rad_motion_reduced} for $u^t$ yields the Larmor power formula \cite{Jackson:1998nia}:
\begin{equation}
	\frac{d E}{d t} = -\frac{2 q^2}{3} a^2\, .
	\label{eq:mot_larmor}
\end{equation}

A calculation of the Newtonian limit of the radiation reaction force experienced by a free-falling charged particle orbiting a Kerr BH can also be found in \cite{Galtsov:1982hwm}. This result can only be recovered if both $\vec{F}_\text{RR}$ and $\vec{F}_\text{tail} ^\phi$ are included in the equations of motion.

In order to visualize the effects of the tail term in the dynamics, we have studied numerically the evolution of particles in the plus configuration,
with and without the inclusion of the tail. Our results, shown in Fig.~\ref{fig:simulations} of Appendix \ref{sec:sim}, illustrate how an unphysical outspiralling of particles occurs when the tail term is neglected.
%
\section{Discussion and conclusions\label{sec:com}	}
We have shown that in order to adequately recover the non-relativistic results in Eqs.~\eqref{eq:mot_lorentz} -- \eqref{eq:mot_larmor} from Eq.~\eqref{eq:rad_DeWitt_Brehme}, in all setups where EM radiation reaction and gravity are relevant, the tail term must \emph{always} be included\footnote{Interestingly, even when this limit is recovered, the conservative part of the tail term, $F^r _\text{tail}$, is still present; it is a completely new term which was not predicted before GR. This term was in fact calculated in the general case, without requiring the Newtonian limit, but instead having the particle be held static, in \cite{Smith:1980tv}. Since it is an intrinsically general relativistic effect, it must be tied to the notion of spacetime curvature. Heuristically, the presence of the BH deforms the electrostatic field lines generated by a point charge in such a way as to mimic the presence of a second charged particle between the original point charge and the BH.}.

Our results disagree with those in Ref.~\cite{2018ApJ...861....2T}. In the latter, the dissipative part of the tail term is neglected altogether, and the equation of motion used to study charged particle dynamics is \eqref{eq:rad_motion_reduced} without the tail term. This implies, in particular, that a particle in a plus configuration orbit will gradually gain energy, given that $\vec{F}_\text{RR}$ will do positive work, as evidenced in \eqref{eq:mot_fforce_rr} and the subsequent discussion. This energy extraction is accompanied by a ``orbital widening" effect. Such a result cannot be correct, as there is no possible source of energy for these particles (in fact, it was this counter-intuitive claim that led us to investigate the tail term). We find that not only such an energy-generating process disappears when the tail is included, but also that it must {\em always} be included, at least in the Newtonian limit.

But why is the tail term and its complicated expression \eqref{eq:rad_tail} so important? Imagine we could indeed neglect it, so the equation of motion after order-reduction would be
\beq
\frac{D u\supmu}{d\tau} &= & \frac{q}{m} F\supmu _{\ \ \nu} u\supnu + \frac{2 q^2}{3 m} \left(\frac{q}{m}\nabla_\alpha F^\mu _{\ \ \nu} u^\alpha u^\nu\right) \nonumber\\
&+& \frac{2 q^2}{3 m} \left(\frac{q^2}{m^2} \left(F^{\mu \nu}F_{\nu \rho} + F^{\nu \alpha}F_{\alpha \rho} u_\nu u^\mu \right)u^\rho \right)\, .
\label{eq:rad_DeWitt_Brehme_2}
\eeq
This equation states that if no EM field is present then the particle will simply follow a geodesic of spacetime, meaning it does not radiate. This would also mean that the gravitational force is fundamentally different from any other, as Eqs.~\eqref{eq:mot_lorentz} -- \eqref{eq:mot_larmor} would have to be corrected so that $\vec{a}$ would actually be the acceleration arising from all forces other than the gravitational force. In GR, gravity is indeed a fundamentally different force; however, it still must make orbiting particles radiate, as their position is varying periodically. This can also be seen from a calculation using the conformal invariance of Maxwell's equations (see Appendix \ref{sec:com_conf}).

To conclude, the take-home message of our work is the following: \emph{To adequately describe the dynamics of a ra\-di\-a\-ting pointlike charge in a vacuum spacetime, the tail term must be included (at least in the Newtonian limit), unless radiation reaction is neglected altogether, or the spacetime curvature can be ignored, in which case a flat space description of radiation, based on the Abraham-Lorentz-Dirac equation, is sufficient.}

In the context of radiating particles around magnetized BHs, previous works have studied the system considered in this work \cite{2018ApJ...861....2T}. The results therein, along with those in Refs.~\cite{Tursunov:2018udx,2021Univ....7..416S}, 
are incorrect, at least in what pertains the ``orbital widening" effect in the Newtonian limit: in fact the authors neglected the tail term, using an argument based only on the conservative part of the latter (which is indeed negligible -- see Eq.~\eqref{eq:mot_compare_conservative}); unfortunately, those works also neglect the dissipative part of the tail term, which plays a crucial role in all relevant cases.

It is highly unlikely that the tail can be ignored in strong field, since it must already be included in the weak field regime. Still, it would be important to perform a full strong field calculation.

\section*{Acknowledgements}
We acknowledge useful feeback from and exchanges with Martin Kološ, Arman Tursunov and Dmitry Gal'tsov.
 V.C.\ is a Villum Investigator and a DNRF Chair, supported by VILLUM Foundation (grant no. VIL37766) and the DNRF Chair program (grant no. DNRF162) by the Danish National Research Foundation. V.C.\ and J.S.S.\ acknowledge financial support provided under the European
Union's H2020 ERC Advanced Grant ``Black holes: gravitational engines of discovery'' grant agreement
no.\ Gravitas–101052587. Views and opinions expressed are however those of the author only and do not necessarily reflect those of the European Union or the European Research Council. Neither the European Union nor the granting authority can be held responsible for them.
This project has received funding from the European Union's Horizon 2020 research and innovation programme under the Marie Sklodowska-Curie grant agreement No 101007855.
We acknowledge financial support provided by FCT/Portugal through grants 
2022.01324.PTDC, PTDC/FIS-AST/7002/2020, UIDB/00099/2020 and UIDB/04459/2020.

\appendix
\section{Simulation of ``plus'' configuration orbits\label{sec:sim}	}
In Sec. \ref{sec:mot_rad} we showed that the inclusion of the dissipative part of the tail term, $F^\phi _\text{tail}$ in \eqref{eq:mot_force_tail}, in the equation of motion is fundamental to adequately capture the dynamics of radiating charged particles in situations where the gravitational force is important when compared to the Lorentz force, see Eq.~\eqref{eq:mot_compare_dissipative}. In particular, we found that if the tail term is neglected for plus configuration orbits then the particle must gain energy as result of the positive work done by $F_\text{RR}$. 

In this appendix, we check our analytical results by performing numerical simulations, where we are not restricted to the adiabatic approximation (i.e. assuming that the orbit is circular with slowly varying parameters). This is still, of course, within the framework of the Newtonian approximation. 

Using Cartesian coordinates $(x,y)$ in the equatorial plane, geometric units, $G=c=1$, and Gaussian units for electromagnetism, $4 \pi \varepsilon_0 = 1$ and $\mu_0 = 4\pi$, the equation of motion is
\begin{equation}
	m \frac{d \vec{v}}{d t} = \vec{F}_\text{N} + \vec{F}_\text{L} + \vec{F}_\text{RR} + \vec{F}_\text{tail}\, ,
    \label{eq:motion}
\end{equation}
where $m$ is the mass of the particle, $\vec{v} = d\vec{x}/dt = v_x \vec{\hat{x}} +v_y \vec{\hat{y}} $ is the velocity, and the various forces are given by \cite{Jackson:1998nia,DeWitt:1964de,Pfenning:2000zf}
\begin{align}
    &\vec{F}_N = 
    - \frac{M m}{r^3} x \vec{\hat{x}} 
    - \frac{M m}{r^3} y \vec{\hat{y}} \, , \\
    &\vec{F}_L = 
    q v_y B_0 \vec{\hat{x}} 
    - q v_x B_0 \vec{\hat{y}} \\
    &\vec{F}_\text{RR} = \frac{2 q^3}{3 m} \frac{v B_0}{r} \left(v_x \vec{\hat{x}} + v_y \vec{\hat{y}} \right) \, , \\
    &\vec{F}_{\text{tail}} = \frac{q^2 M}{r^3} \Bigg(
    \lr{\frac{x}{r} - \frac{2}{3} v_x }\vec{\hat{x}} 
    + \lr{\frac{y}{r} - \frac{2}{3} v_y }\vec{\hat{y}} \Bigg) \, .
\end{align}
Here $M$ is the BH mass, $r=\sqrt{x^2 + y^2}$ is the radial coordinate, $q$ is the charge of the particle and $B_0$ is the magnetic field intensity. 

\begin{figure*}[!t]
    \centering
    \includegraphics[width= .45 \textwidth]{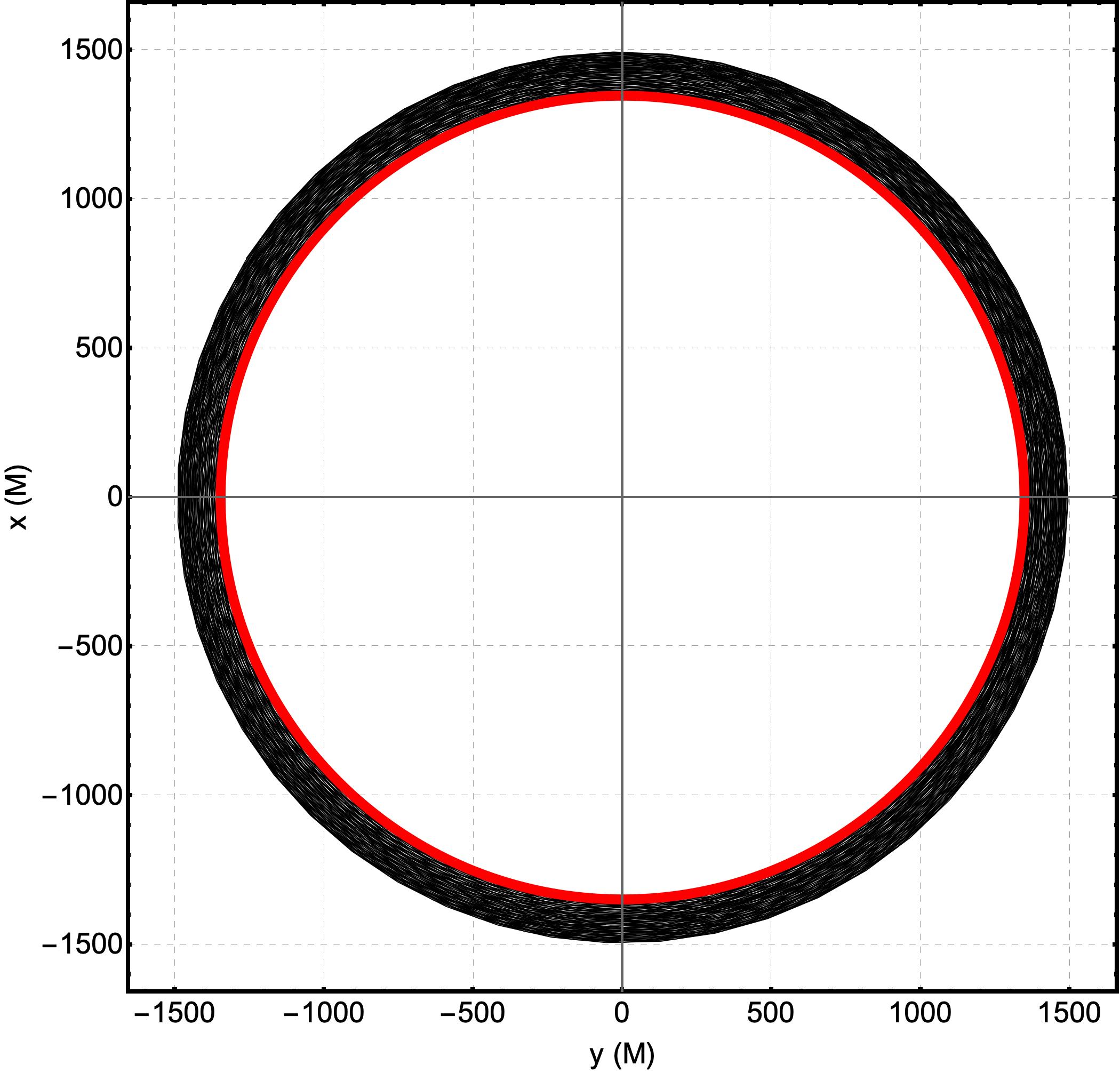}
    \hspace{1cm}
    \includegraphics[width= .46 \textwidth]{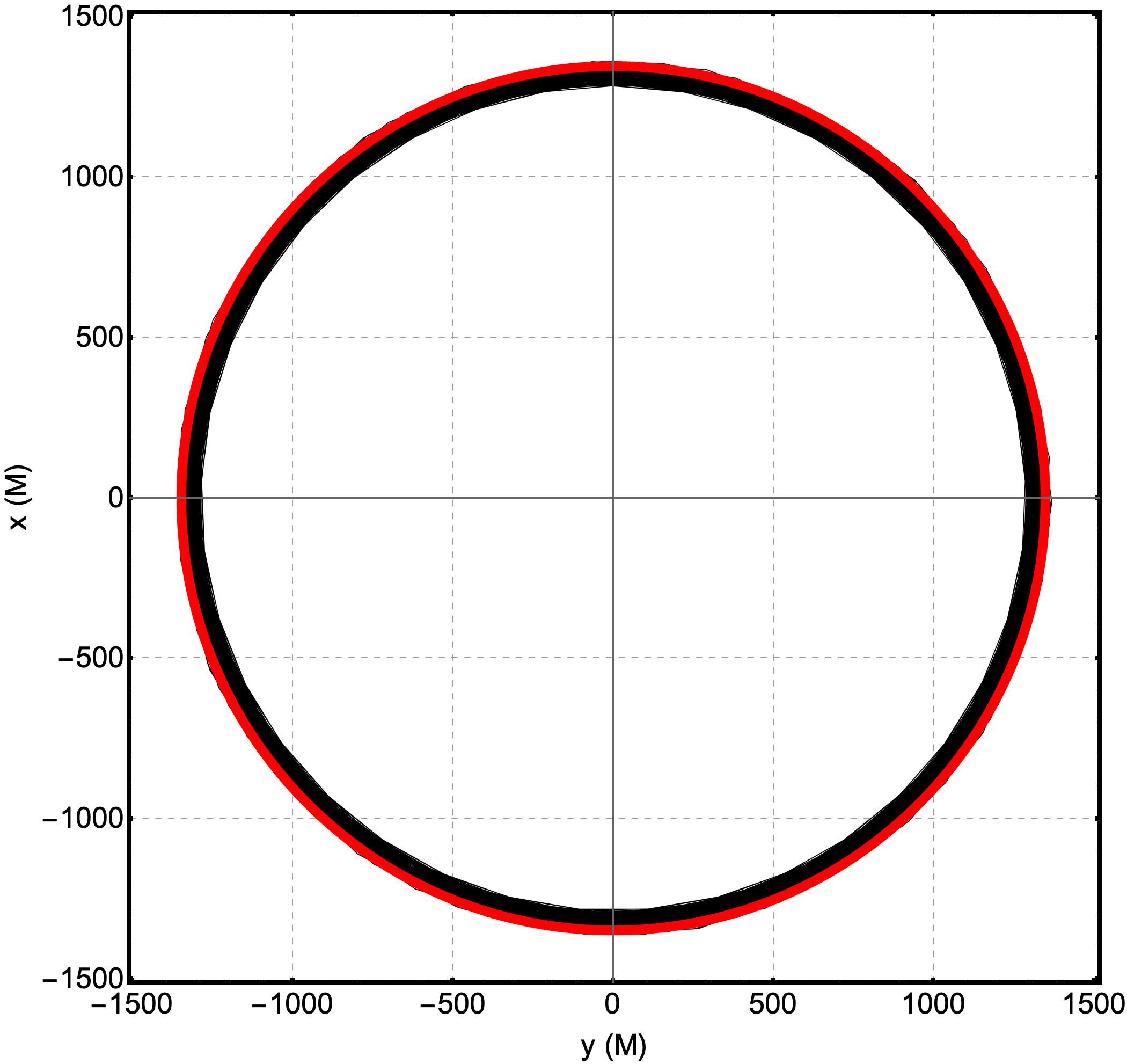}
   \caption{Trajectories of a charged particle in a plus configuration orbit, with the parameters given in Table~\ref{tab:simulation}. The red circle represents the initial orbit, with the subsequent particle motion plotted in black. \emph{Left panel} -- motion without the inclusion of $\vec{F}_\text{tail}$. \emph{Right panel} -- motion with the inclusion of $\vec{F}_\text{tail}$. We can see that if the tail is neglected then the particle gains energy and the orbit drifts away from the BH, as a consequence of the positive work done by $\vec{F}_{RR}$ (left). By contrast, if the tail term $\vec{F}_\text{tail}$ is included then it does enough negative work so that the particle loses energy and starts falling into the BH (right).}
    \label{fig:simulations}
\end{figure*}

These equations were used to numerically simulate the motion of particles in plus configuration orbits, using the parameters given in Table~\ref{tab:simulation}. In order to illustrate the importance of the tail term, we present two simulations, with and without including the term $\vec{F}_\text{tail}$ in the equations of motion (Fig.~\ref{fig:simulations}). Since $\vec{F}_{RR}$ is doing positive work, when the tail is excluded the particle starts drifting away from the BH, as a result of gaining energy. By opposition, as we expected from our analytical results, the particle starts falling into the BH when the tail term is included. 
\begin{table}[ht]
    \centering
    \begin{tabular}{c|c|c}
       \hline
       Qty & Value & Simulation \\
       \hline
        $M$ & $10 \, M_\circledcirc$ & $1$ \\
        $B_0$ & $10^8 \, G$ & $1.20 \times 10^{-11}$ \\
        $q$ & $7\times 10 ^{15} \, C$ & $1.44 \times 10^{-5}$ \\
        $m$ & $5\times10^{-12} \, kg$ & $5 \times 10^{-12}$ \\
        $r_0$ & $2 \times 10^7 \, m$ & $1.35 \times 10^3$ \\
        \hline
    \end{tabular}
    \caption{Parameters used in our simulations. Second column gives values in SI units and solar masses ($M_\circledcirc$) while the third gives the values used in the actual simulation (after normalizing to BH mass).}
    \label{tab:simulation}
\end{table}

This clearly illustrates that the orbit widening effects reported in \cite{2018ApJ...861....2T, Tursunov:2018udx} result from incorrectly excluding the tail term from the equation motion, at least in the Newtonian limit. 
%
\section{Radiation reaction in the Newtonian limit\label{sec:com_conf}	}
Here we use a simple setup to establish that an orbiting charged particle does radiate according to Larmor's power formula in the Newtonian limit. Consider the (non-physical) metric
\begin{equation}
g=e^{2\Phi} \eta = e^{2\Phi} \left( - dt^2 + dx^2 + dy^2 + dz^2 \right)\,,\\
\end{equation}
\begin{equation}
\Phi = -\frac{M}{r} \,,
\end{equation}
which, under the assumptions of Eq.~\eqref{eq:mot_assumptions}, has the appropriate Newtonian limit \eqref{eq:mot_newtonian_limit}. Since this metric is conformal to the Minkowski metric, and the Maxwell equations are conformally invariant in $(3+1)$ dimensions, there will be no tail. However, as it is not a solution of the vacuum Einstein field equations, the Ricci tensor will not vanish. In fact, it is given by (see \cite{Wald:1984rg}, appendix D)
\begin{equation}
R_{\mu\nu} = -2\partial_\mu\partial_\nu\Phi - \Box \Phi \, \eta_{\mu\nu} + 2\partial_\mu\Phi\partial_\nu\Phi - 2(\partial_\alpha\Phi\,\partial^\alpha\Phi) \eta_{\mu\nu} \, .\nonumber
\end{equation}
The radiation reaction force term corresponding to the Ricci tensor in Eq.~\eqref{eq:rad_DeWitt_Brehme} is
\begin{equation}
F^\alpha = \frac{q^2}3 (g^{\alpha\mu}+u^\alpha u^\mu) R_{\mu\nu} u^\nu \, .
\end{equation}
Because of the orthogonal projector along $u^\alpha$, we can drop all terms in $R_{\mu\nu}$ proportional to $\eta_{\mu\nu}=e^{-2\Phi}g_{\mu\nu}$. For circular orbits, $u^\nu \partial_\nu \Phi = 0$, and so the third term in $R_{\mu\nu}$ can also be dropped. Therefore we obtain
\begin{equation}
F^\alpha = - \frac{2q^2}3 (g^{\alpha\mu}+u^\alpha u^\mu) (\partial_\mu\partial_\nu\Phi) u^\nu \, ,
\end{equation}
and so, since $\Phi$ is time-independent,
\begin{equation}
F^0 = - \frac{2q^2}3 u^0 u^\mu u^\nu (\partial_\mu\partial_\nu\Phi) \, .
\end{equation}
Assuming that the circular orbit is on the plane $z=0$, and also that the particle is momentarily at $(x,y)=(R,0)$ in this plane, we have, for large $R$,
\begin{equation}
u \simeq \frac{\partial}{\partial t} + v \frac{\partial}{\partial y} \, , \qquad v=\sqrt{\frac{M}{R}} \, .
\end{equation}
We then obtain
\begin{equation}
F^0 \simeq - \frac{2q^2}3 v^2 \frac{\partial^2 \Phi}{\partial y^2} = - \frac{2q^2}3 v^2 \frac{M}{R^3} \, ,
\end{equation}
that is,
\begin{equation}
F^0 \simeq - \frac{2q^2}3 \frac{M^2}{r^4} = - \frac{2q^2a^2}3 \, ,
\end{equation}
where $a$ is the Newtonian acceleration of the orbit, in accordance with Larmor's formula \eqref{eq:mot_larmor}.

Thus, even in GR, we find that a particle under the action of a gravitational field radiates according to Larmor's formula when the Newtonian limit applies. In this particular case, because of the conformal invariance of Maxwell's equations, the radiation reaction force comes from the Ricci tensor terms in \eqref{eq:rad_DeWitt_Brehme}. However, when a vacuum metric is considered, those terms are identically zero, and so the radiation reaction can only come from the tail, since the terms $\propto \frac{D^2 u\supmu}{d \tau^2}$ vanish identically on a geodesic.
%

\end{document}